\documentclass[preprint,showpacs,titlepage,aps,prd,tightenlines,amsmath,byrevtex,nofootinbib]{revtex4}
\def\lsim{\ ^<\llap{$_\sim$}\ }

\usepackage[dvips]{color}
\usepackage{epsfig}
\begin{document}
\baselineskip 18pt
\def\today{\ifcase\month\or
 January\or February\or March\or April\or May\or June\or
 July\or August\or September\or October\or November\or December\fi
 \space\number\day, \number\year}
\def\thebibliography#1{\section*{References\markboth
 {References}{References}}\list
 {[\arabic{enumi}]}{\settowidth\labelwidth{[#1]}
 \leftmargin\labelwidth
 \advance\leftmargin\labelsep
 \usecounter{enumi}}
 \def\newblock{\hskip .11em plus .33em minus .07em}
 \sloppy
 \sfcode`\.=1000\relax}
\let\endthebibliography=\endlist
\begin{flushright}
NORDITA-2008-7
\end{flushright}
\title{Learning more about what can be concluded from the observation of 
neutrinos from a galactic supernova}
\vskip 1.5 true cm
\renewcommand{\thefootnote}
{\fnsymbol{footnote}}
\author{Solveig Skadhauge\email{E-mail: solveig@nordita.org}}
\affiliation{ Nordita, AlbaNova University Center, Roslagstullbacken 23, 
SE-10691 Stockholm, Sweden} 
\author{Renata Zukanovich Funchal\email{E-mail: zukanov@if.usp.br}}
\affiliation{Instituto de F\'{\i}sica, Universidade de S\~{a}o Paulo, 
 C.\ P.\ 66.318, 05315-970 S\~{a}o Paulo, SP, Brazil}
\vskip 1.5 true cm
\begin{abstract}
  We investigate what one can hope to learn about the parameters that
  describe the neutrino fluxes emitted by the explosion of a galactic
  supernova using the observations of a megaton-size Water-Cherenkov
  detector.  We calculate the allowed regions that can be obtained by
  fitting these parameters to a simulated observation of events by
  such a detector. All four available detection channels (inverse beta
  decay, charge and neutral current on oxygen and elastic scattering
  on electrons) are included in the fit and we use a ten dimensional
  parameters space. Nine parameters describe the initial neutrino
  fluxes and are referred to as the supernova parameters. Furthermore, 
  we include the dependence on the Chooz mixing angle $\theta_{13}$, which
  controls the matter effects that the neutrino undergoes in the
  outer-parts of the supernova.   
  If we do not make any extra assumption on these parameters, we 
  show that one can hope to determine $\theta_{13}$ quite well 
  whereas, except for the parameters describing the $\bar\nu_e$ flux,  
  most of the supernova parameters are rather difficult to 
  constrain, even if the four detection channels could be completely 
  separated.  
\end{abstract}
\maketitle
\section{\label{sec:intro} Introduction}
The phenomenon of a core-collapse Supernova (SN) explosion embraces 
vastly different areas of physics and involves all the known 
fundamental interactions~\cite{snreview}.  
The core-collapse itself is driven by gravitational effects;  
the thermodynamics is controlled by electromagnetic and strong forces  
and naturally the weak force plays a major role in the energy loss 
through emission of neutrinos. 
Evidently, the study of supernovas is a promising playground for 
testing new physics and to learn about particle properties. 

Among the particles one hopes to learn more about are the neutrinos.
Indeed the major part of the enormous energy liberated in a
core-collapse (type-II) supernova is emitted in the form of
neutrinos. Naturally, one might also attempt to use our knowledge of
neutrino properties to extract information about the supernova physics
from observations of the neutrino fluxes. In particular, as the
neutrinos are emitted from the interior of the star, they offer a 
way to probe the physics of the core-collapse.  But independent
of whether one attempts to learn about supernova physics or about
neutrino properties from a SN observation, ultimately, when comparing
data with theory one needs to take into account the uncertainties on
the supernova parameters as well as on the neutrino parameters.

Many efforts have been made in order to understand the complex physics
involved and to predict the emitted neutrino
fluxes~\cite{garching,livermore}.  When discussing supernova neutrino
there are only three distinguishable flavors, as the $\mu$ and $\tau$
neutrinos, as well as their antineutrinos, have identical properties.
We will denote the $\mu$ and $\tau$ neutrinos and antineutrinos by $\nu_x$.
Thus $\nu_x$ along with the electron neutrino ($\nu_e$) and the
electron antineutrino ($\bar \nu_e$) constitute the three supernova
neutrino species.  The neutrinos from the core-collapse supernova are
expected to be almost thermal, since they are in fact trapped inside
what is known as the neutrino-sphere. Besides the very early universe
a supernova is probably the only place neutrinos thermalize. However,
small deviations from the thermal spectrum are expected and how narrow
or wide the energy spectra will be is in general described by a
pinching parameter.  Therefore, it is a good approximation to
parametrize each initial neutrino spectrum by three parameters; the
average neutrino energy, the total emitted energy (or integrated
luminosity) and the pinching parameter.

During the last few years evidences of neutrino masses and mixing have
been gathered and today most of the neutrino oscillation parameters
have been determined with good precision, from the observation of
solar and atmospheric neutrinos as well as other terrestrial neutrino
experiments~\cite{solar,atm,kamland,k2k,MINOS}.  Some important
parameters remain unknown.  The absolute neutrino mass scale, $m_0$,
is only bounded from above by cosmological data $m_0 \lsim 0.2-0.7$
eV~\cite{wmap}. The same is true for the mixing angle $\theta_{13}$,
$\sin^2\theta_{13} \lsim 0.04 $~\cite{chooz}.  Furthermore, the
pattern of neutrino masses is not established yet: we do not know if
nature has chosen the normal ($m_3 > m_2 > m_1$) or the inverted ($m_2
> m_1 > m_3$) mass hierarchy, where $m_1$ ($m_3$) is the mass of the
neutrino state most (least) populated by the $\nu_e$ component.
Finally, we do not know anything about the existence of CP violating
phases in the neutrino sector~\footnote{CP-violation effects are
ignored in our analysis. It was explicitly shown in
\cite{balantekin} that there is no net CP effect due to standard
neutrino oscillations in a SN core-collapse.}.

In particular, for supernova neutrinos the neutrino mass hierarchy and
the Chooz angle, $\theta_{13}$, are very important, as the so-called 
Mikeyev-Smirnov-Wolfenstein (MSW) H resonance strongly depends on 
them~\cite{mswsn,Minakata:2000rx}.
Recently, there has been several investigations of the possible impact
of the neutrino-neutrino interactions in the dense neutrino region
inside the neutrino-sphere in the supernova~\cite{collective}.  These
self interactions may cause collective effects and thus influence the
neutrino survival probability as a function of energy. Under realistic
supernova density profiles the collective effects are negligible for
the normal mass hierarchy.  In the case of the inverted hierarchy it
seems that a swapping of the $\nu_e$ and $\nu_x$ energy spectra above
a certain critical energy occurs, and at the same time a swapping of
the antineutrino spectra occurs. In the latter case there are
indications that the spectra might be smeared out and therefore does
not exhibit a sharp interchange of the spectra at a certain energy. In
principle, these swapped spectra should then be used as input when
calculating the MSW effects in the outer parts of the
supernova. However, since the impact of the neutrino-neutrino self
interactions is still being debated, we will in this paper only
consider the normal mass hierarchy and we can thus neglect the
collective neutrino-neutrino effects.

One can imagine two very different sources of supernova neutrino
signals.  The diffuse supernova neutrino background (DSNB), arising
from all past supernova explosions; and the {\em lucky} event of a
galactic supernova. In the latter case one expect about $10^4 - 10^5$
neutrino events in a megaton scale detector in a 10 seconds time
interval.  Therefore, this is practically free of background. On the
other hand, background is the major issue for the DSNB detection and 
one expect only to be able to extract a signal in a rather small energy 
window and most likely only for the dominant detection channel. Therefore,
depending on the detector, there will be sensitivity only to the
electron antineutrino flux (Water-Cherenkov and Scintillator) or only
to the electron neutrino flux (Liquid Argon).  On the contrary, for 
galactic SN observations one expect to have several detection channels
available, each with sensitivity to different neutrino flavor
compositions - a feature which is crucial for the pinning down of the
neutrino parameters.  As is well-known, a galactic supernova explosion
is a rare event.  About two per century is the best we can hope for.
The optical observation is actually likely to be obscured by dust
etc., and in this light it seems even more crucial to get a better 
understanding of the SN neutrino fluxes.

Here we will focus on the investigation of what can be learned from
an observation of the neutrino burst from a galactic supernova. 
We will furthermore restrict ourself to what can be measured by a
Water-Cherenkov detector.  We aim at figuring out what can be learned
about supernova physics as well as what can be learned about the
neutrino parameters simultaneously.  We thus continue previous
studies~\cite{sn1,oldfit,mina}, but we will increase the parameter
space, taking into account all important parameters in what
optimistically could be referred to as a {\em full parameter space}
fit. In the previous studies only the dominant detection channel has
been considered or some of the parameters have been fixed. In most
earlier works the pinching parameters were fixed, but in
Ref.\cite{mina} it was shown that, when considering only the dominant
inverse beta decay detection channel, these constitute important
uncertainty factors when attempting to extract information about the
supernova parameters.

In fact, the authors of Ref.~\cite{mina} have discovered a 
degeneracy,  which only appears when including the pinching parameters, 
between the mean energy $\langle E_x \rangle$, the luminosity $L_x$ and 
the pinching parameter $\beta_x$ of $\nu_x$ flux.
By degeneracy we mean that, for a given allowed point in  
($\langle E_x \rangle$, $L_x$, $\beta_x$), i.e. a point 
where the data are explained (one can always take the input values 
themself), it is possible to construct a different allowed point 
which has a larger value of $\langle E_x \rangle$ by 
increasing $\beta_x$ and decreasing the luminosity $L_x$. 
As this degeneracy covers almost the complete parameter space, 
the determination of the correct value of 
$\langle E_x \rangle$ (as well as $\beta_x$ and $L_x$) becomes very hard. 
Indeed the allowed region for $\langle E_x \rangle$ 
was shown to include the whole range from 15 MeV to 30 MeV 
as expected from supernova simulations~\cite{mina}. 
This degeneracy occurs since shifting the parameters in the 
mentioned fashion, maintains the $\bar \nu_e$ energy-spectrum at the 
Earth almost identical, with only small differences in the 
low energy spectrum. Due to the threshold (which is about 5 MeV) on the 
inverse-beta decay channel, the Water-Cherenkov detector 
is not sensitive to the spectrum at low energies.
Clearly the $\bar \nu_e$ parameters have also to be adjusted correspondingly, 
but as the allowed regions are still rather small, we will not speak 
of a degeneracy in these variables. 
However, only the inverse beta decay channel 
has been considered in Ref.~\cite{mina} and it is not 
clear to what extend the inclusion of the other detection 
channels might break down this degeneracy. 
Indeed in Ref.\cite{sn1} several degeneracies in the 
only-inverse-beta-decay case were found to be broken down when 
considering all four detection channels. 
This is one of the motivations for the present work. 

In our {\em full parameter space} fit, we use a
total of ten fitting parameters. Nine parameters describe the initial
supernova neutrino flux ($\nu_e$, $\bar \nu_e$ and $\nu_x$
luminosities, mean energies and pinching parameters), as stated above
three parameters are needed for each of the three neutrino species.
Furthermore, we also freely vary the most important neutrino
parameter, the Chooz angle $\theta_{13}$.  However, as explained above
we fix another important neutrino parameter, the neutrino mass
hierarchy, as normal.  The terminology {\em full parameter space}
should therefore, as usual, be taken with a grain of salt.  There
might be a number of other parameters, such as deviation of the
supernova density profile from the assumed $\rho^{-3}$ form~\cite{dens},
effects from sterile neutrinos~\cite{Keranen:2007ga} or from new 
interactions~\cite{balantekin}, deviation from the assumed 
spectral forms and yet undiscovered effects that may influence 
the flux of neutrinos from a supernova.

\section{\label{sec:analysis} The analysis procedure and the parameter  space coverage}
In this section we describe the details of our analysis procedure
and the chosen parameter space. As mentioned in the introduction we
will investigate how well the supernova and neutrino parameters can be
determined by the observation of neutrinos from a galactic SN by a
future megaton scale Water-Cherenkov detector.  We will take into
account all major parameters (with exception of the neutrino mass
hierarchy, here assumed to be normal) that can influence the neutrino
flux, thus expanding the parameter space as compared to earlier
works. We vary a total of 10 parameters to be fitted by data: the
$\nu_e$, $\nu_{\bar e}$ and $\nu_x$ luminosities ($L_e$, $L_{\bar e}$
and $L_x$), mean energies ($\langle E_e \rangle$, $\langle E_{\bar e}
\rangle$ and $\langle E_x \rangle$), pinching factors ($\beta_e$,
$\beta_{\bar e}$ and $\beta_x$), 9 SN parameters, and a single
neutrino quantity, $\sin^2 \theta_{13}$.
 
We will assume that the initial supernova neutrino fluxes emitted at
the respective neutrino-spheres can be parametrized with the spectrum
as suggested in Ref.\cite{Mirizzi:2005tg}. Thus, for each neutrino
specie $i=e, \bar e, x$, we assume the energy spectrum to be of the form
\begin{equation}
 \phi_i^0=\frac{\beta_i^{\beta_i+1}}
 {\langle E_i \rangle ^2 \, \Gamma (\beta_i+1)} \, 
  \left( \frac{E}{\langle E_i \rangle} \right) ^{\beta_i-1} 
 \exp ({-\beta_i \frac{E}{\langle E_i \rangle}}) 
  \; \; ,
\end{equation}
where $\beta_i$ is the pinching parameter, $E$ the neutrino energy and
$\langle E_i \rangle$ the $\nu_i$ mean energy.  For $\beta_i \ge 3$
the spectrum is pinched with suppressed low and high energy
tails, whereas for $\beta_i \le 3$ the spectrum is anti-pinched
(broader). The $\nu_x$ neutrinos decouple at a smaller radius and will
therefore be hotter.  Due to different charge-current interaction also
the electron antineutrinos will decouple before the electron
neutrinos. Correspondingly, a hierarchy of the form $\langle E_e
\rangle \le \langle E_{\bar e} \rangle \le \langle E_x \rangle$ is
expected.

The unoscillated flux at distance $D$ from the SN is given by
\begin{equation}
  F^0_{\nu_i} = \frac{L_i}{4 \pi D^2} \, \phi^0_i(E)\, ,
\end{equation}
and we will fix the distance to 10 kpc. 
The neutrinos which are emitted from the interior of the star may 
undergo various flavor transitions due to MSW matter  effects, 
when passing through the outer layers of the supernova.  
For the normal hierarchy, the the $\nu_e$ and 
$\bar \nu_e$ survival probabilities, $P_{ee}$ and 
$P_{\bar e \bar e}$, respectively, are approximated by
\begin{eqnarray}
P_{ee}&\simeq & P_H |U_{e2}|^2 + (1 - P_H) |U_{e3}|^2 , \label{pue3nh} \\
P_{\bar e\bar e} & \simeq & |U_{e1}|^2  \label{pue3nh2}  \;,
\end{eqnarray}
where $U_{\alpha,i}\;,\alpha=e,\mu,\tau\; ,i=1,2,3$ are the
Maki-Nakagawa-Sakata neutrino mixing matrix elements and we have used
here the standard parameterization. We have fixed the value of the
solar mixing angle $\theta_{12}=0.575$ rad ($\sin^2 \theta_{12}=0.3$)
and the atmospheric mass squared difference $\Delta m_{31}^2$ is set 
to be $3 \times 10^{-3}$ eV$^2$, as their variation have little impact
on the final neutrino fluxes. We will disregard Earth matter effects
and therefore the exact value of solar mass squared difference is
irrelevant. Also the value of the atmospheric mixing angle is not
relevant as the muon and tau neutrinos are indistinguishable. $P_H$ is
the hopping probability that can be written as

\begin{equation}
  P_H = \exp \left[ -  \sin^2\theta_{13}
  \left( \frac{1.08 \cdot 10^{7}}{E} \right)^{2/3}
  \left( \frac{|\Delta m_{31}^2|}{10^{-3}} \right)^{2/3}  4^{1/3} \right]
   \;.
\label{eq:ph}
\end{equation}

The final fluxes arriving at Earth are simply given by
\begin{eqnarray}
 F_{\nu_e} &=& F^{0}_{\nu_e} \, P_{ee} + F^{0}_{\nu_x}\, (1-P_{ee}), \\
 F_{\bar\nu_e} &=& F^0_{\bar\nu_e} \, P_{\bar e \bar e}
 + F^0_{\bar \nu_x} \, (1-P_{\bar e\bar e}), \\
 F_{\nu_\mu} + F_{\nu_\tau} &=&  F^{0}_{\nu_e} \, (1-P_{ee})
+ F^{0}_{\nu_x} \, (1+P_{ee}), \\
 F_{\bar \nu_\mu}+ F_{\bar \nu_\tau} &=&
 F^0_{\bar \nu_e} \, (1- P_{\bar e \bar e})
 + F^0_{\bar \nu_x} \, (1+ P_{\bar e \bar e})\, .
\label{snfluxfinal}
\end{eqnarray}
These neutrino fluxes depend on our 10 dimensional parameters space, 
which is given by
\begin{eqnarray}
&&  \langle E_i \rangle \; , \; 
  L_i \; , \; 
  \beta_i \; , \;\;\;\; i= e, \bar e, x  \\
&&  \sin^2 \theta_{13}
\label{paramspace}
\end{eqnarray}
and we refer to the first 9 parameters as the SN parameters.

There are four known detection channels for  
a water-Cherenkov detector: the dominant inverse beta decay (IB); 
the charge current on oxygen (CC-O); the neutral current on oxygen 
(NC-O) and the elastic scattering on electrons (ELAS). 
For a detailed description of these four possible channels in a 
Water-Cherenkov detector, please see \cite{sn1}.

We will analyze three different cases for fitting the supernova 
and neutrino parameters. 
\begin{itemize}
\item {\bf Case A:} We assume that only the inverse beta decay channel is 
available. In this case there is no dependence on the $\nu_e$ 
parameters. Moreover, since we consider only the normal hierarchy the 
allowed parameter space will even be independent of the 
Chooz angle ($\theta_{13}$), since $P_{\bar e \bar e}$ is 
constant (see eg. Fig.1 of Ref.\cite{sn1}).  
Therefore, in this case only 6 parameters are left to be determined by data. 
\item {\bf Case B:} Here we will assume that there are four available 
detection channels and the detected neutrino fluxes are sensitive to all 
10 parameters. 
\item {\bf Case C:} In this case we also assume that all four detection 
channels are available, but we impose the constraint $L_e= L_{\bar e}$, 
leaving us with 9 free parameters. 
\end{itemize}
The enormous flux of neutrinos which will arrived at the Earth from a
galactic supernova, makes case A a highly pessimistic scenario, as
it seems very likely that it will be possible to measure and also
separate at least some of the other detection channels. Our main
motivation for including this case is in order to be able to compare
to previous works.  Similarly, the condition $L_e = L_{\bar e}$ has
been used in previous works, so we include this case for easy
comparison.  We would like to note that we assume that all the four
channels can be completely separated, like it was done in
Ref.\cite{sn1}.  This is certainly not a realistic assumption but it
allows us to establish what would be the best attainable results for a
future Water-Cherenkov detector.  We should however point out that
with the addition of gadolinium~\cite{gadolinium} it should be at
least possible to do a fairly good separation by using the known
directional forms of each event type.

We simulate the signals for a given set of input values of the 10 
parameters of Eq.~(\ref{paramspace}) and try to find the allowed 
regions of these parameters minimizing a $\chi^2$ function. For this 
purpose we define the simple $\chi^2$ function 
\begin{equation}
  \chi^2 = \sum_{i=1}^{N_{\rm bin}} \frac{(N_i^{\rm th}-N_i^{\rm obs})^2}{N_i^{\rm obs}}
  \;,
\end{equation}
where $N_i^{\rm obs}$ and $N_i^{\rm th}$ are, respectively, the
simulated and fitted number of events in the $i$-th energy bin, and we
take the number of bins, $N_{\rm bin}$, to be 40. All bins have a
width of 2.5 MeV and we set the threshold at 5 MeV. For case A, the
$\chi^2$ function only includes events from the inverse beta decay
channel;
\begin{equation}
 \chi^2_{\rm Case \; A}= \chi^2_{\rm IB}\, , 
\end{equation} 
whereas we use all four available channels for cases B and C
\begin{equation}
 \chi^2_{\rm Case \; B,\; Case \; C}= \chi^2_{\rm IB}  + \chi^2_{\rm CC-O} 
+\chi^2_{\rm NC-O} + \chi^2_{\rm ELAS}\;, 
\end{equation} 
and we calculate the 3 $\sigma$ allowed areas in the projected two dimensional 
space using 2 degrees of freedom.

Naturally the allowed areas will depend on the assumed true values of
the parameters, what we will refer to as the input parameters. We will
just look at one illustrative example and for that we have chosen the
input values $\langle E_e \rangle = 12 \; {\rm MeV} \;,\; \langle
E_{\bar e} \rangle = 15 \; {\rm MeV} \;,\; \langle E_x \rangle = 18 \;
{\rm MeV} $ and furthermore we take all the integrated luminosities to
be equal; $L_i=0.5 \times 10^{32}$ Ergs. The spectral indices are
assumed to be $\beta_e = 5$, $\beta_{\bar e}=5$ and $\beta_x=4$. Finally, 
the neutrino parameter $\sin^2\theta_{13}= 10^{-6}$.  These values are
all chosen equal to the input values of Ref.\cite{mina} as this allows
for an easier comparison. For the detector size we have also taken the
same value as in Ref.\cite{mina} with a fiducial volume of 720 kton.
This value is somewhat high but it should be remembered that the
fiducial volume for a galactic supernova detection is larger than that
for eg.  atmospheric neutrino detection and therefore this value is
not unrealistic.

We will use a MINUIT~\cite{minuit} based code for finding the
3$\sigma$ allowed regions.  A number of constraints on the parameters
have also been implemented;
\begin{eqnarray}
&&  5 \leq \langle E_e \rangle/{\rm MeV} \leq 17\, ,  \\
&&  5 \leq \langle E_i \rangle/{\rm MeV} \leq 35  \;\;,\;i=\bar e,x  \\
&&  0 \leq \beta_i \leq 25,   \;\;,\;i= e,\bar e, x  
\end{eqnarray}
such as to avoid extremely unrealistic values.  
Also, for the case A, the ratio between the neutrino fluxes are 
not allowed to  become too large. Especially this means that 
we take the $\chi^2$ minimum value 
to be the minimum which is the local minimum closest to the input values. 
To be specific, we set all parameters to their input values and let MINUIT 
fall into the local minimum from there.

\section{\label{sec:discussion} Discussion}
In this section we will present our results, and also compare them  
to previous works. Furthermore, we investigate ways to 
break down the degeneracies which exists in the parameter space.

\begin{figure}[htp]
\centering
\includegraphics[width=14.5cm,height=18.cm]{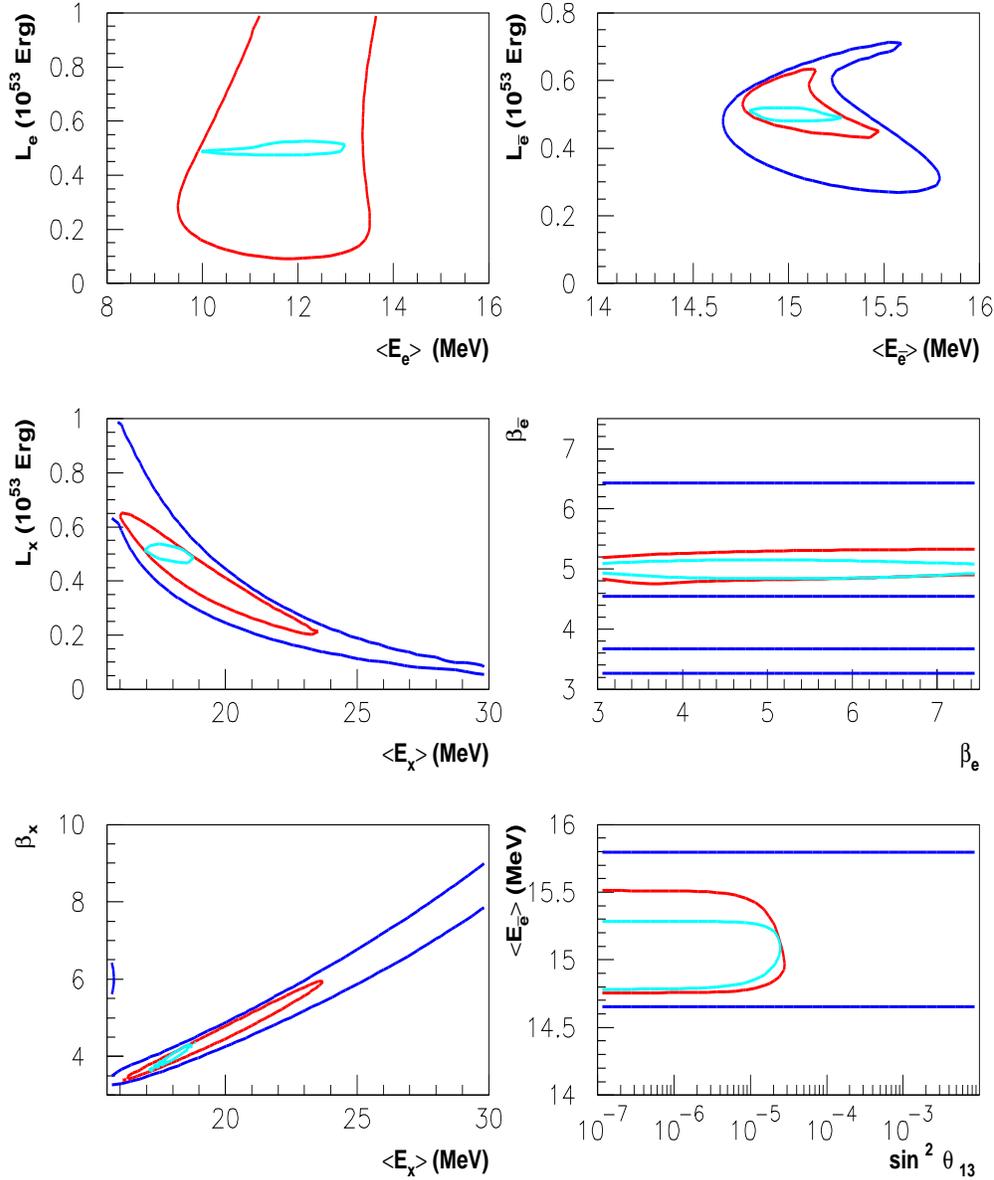}
\vspace{0.1cm}
\caption{The 3 $\sigma$ allowed regions for the three cases A, B and C 
as explained in section \ref{sec:analysis}. 
The outermost dark (blue) contour assumes that only the inverse 
beta decay channel is available, and thus only depends on 6 variables 
(case A). The middle dark (red) contour assumes that all 4 channels are 
available and in this case there is in total 10 free variables (case B). 
For the light (cyan) contour we also assume that all 4 detection 
channels are available, 
but we implement the condition $L_e=L_{\bar e}$, thus leaving only 
9 free variables (case C). For the case A, there is no dependence on the 
electron-neutrino parameters and therefore this case does not constrain 
any of the variables in frame 1.}
\label{fig:cont}
\end{figure}
In Fig.\ref{fig:cont} we show the 3 $\sigma$ allowed areas 
for the three cases A, B and C in six different planes of
two parameter projections. It is seen that the allowed 
regions are quite different in each case. 
The allowed regions for case A are the largest 
and the ones for case C are the smallest, which is clear 
from the definitions of the three cases. 
The only exception is for $\beta_e$ which remains completely 
undetermined even in case C.

By looking at the contours for case A, we see that we overall agree
with the findings of Ref.\cite{mina}.  Indeed the degeneracy in the
$\langle E_x \rangle$ - $\beta_x$ - $L_x$ parameters is also clearly
exhibited in Fig.\ref{fig:cont} (panels 3 and 5).  Although, there are
minor differences in the size of the areas, the shapes of the allowed
areas are the same as in Ref.\cite{mina}.  The small differences must
be due to slightly different analysis procedures. For instance, the
width of the energy bins are different in the two analysis.  As
already explained there is no sensitivity to the $\nu_e$
parameters as well as to $\sin ^2 \theta_{13}$ in case A.  The
determination of the $\bar \nu_e$ parameters are not impressive but
still fairly good. The uncertainty on the value of $\langle E_{\bar e}
\rangle$ is about 5 \%, $L_{\bar e}$ is determined within 50\% and 
$\beta_{\bar e}$ is consistent with values in two distinct island, 
one including the input value.  Clearly the most important and
problematic fact is that the $\nu_x$ parameters are left undetermined.
The degeneracy extends over the whole realistic range of parameter. 
Therefore, without further input, it is unfortunately
impossible to say something conclusive about the initial $\nu_x$ flux.
 
\begin{figure}[htp]
\centering
\includegraphics[width=8cm,height=8.cm]{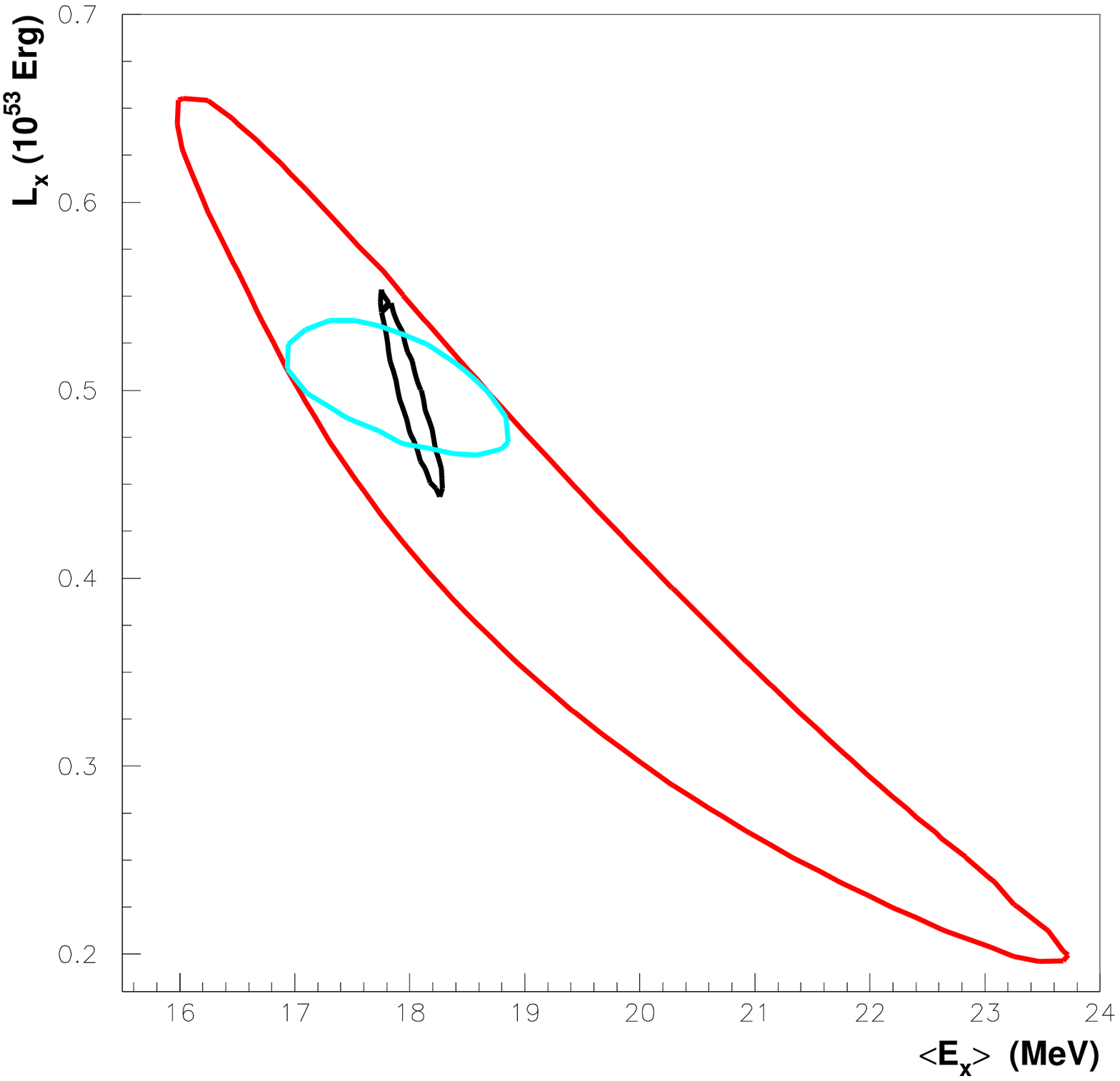}
\includegraphics[width=8cm,height=8.cm]{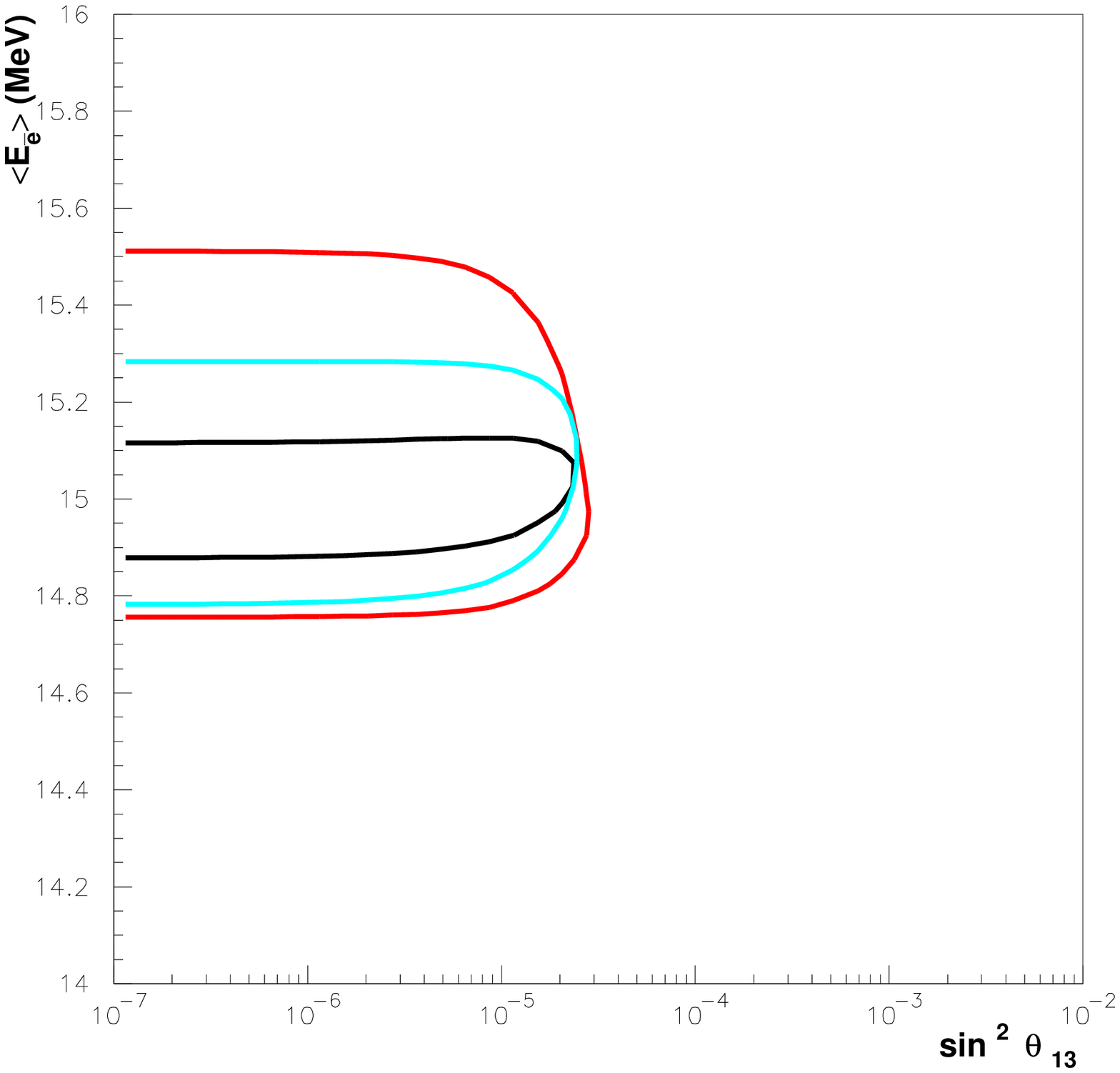}
\vspace{0.1cm}
\caption{The innermost dark (black) contour  shows the 
3$\sigma$ allowed region when fixing the $\beta_i$'s at 
their input values. We assume that all four channels are 
available. 
The outermost dark (red) and the light (cyan) contours are 
identical to the ones in Fig.\ref{fig:cont} and shown for 
easy comparison. 
}
\label{fig:etafixed}
\end{figure}

In case B we find rather large allowed regions for some of the
parameters, but considerably smaller than for case A.  Although the
degeneracy is partially broken, there is still correlation among
the $\nu_x$ parameters so a large elongated area is still seen in
panels 3 and 5 of Fig.\ref{fig:cont}.  We believe that the main reason
for the fact that much of the degenerate area survives is simply that
the other detection channels (CC-0, NC-0 and ELAS) have a much smaller
number of events. For an idea of the orders of magnitudes of each
event type, please see Table II in Ref.\cite{sn1}.  By comparing with
the results of Ref.\cite{sn1} it becomes clear that the inclusion of
the pinching parameters have a very large impact on the size of the
allowed areas for the average energies, in particular that of the
$\nu_x$ neutrinos.  This is also illustrated in
Fig.\ref{fig:etafixed}.  The fact that, to a large extend the
degeneracy is still left even after including all four detection
channels, should be viewed as a large obstacle for the determination
of the $\nu_x$ parameters.  Below we will study some possible ways of
overcoming this degeneracy.

For case B it is seen that the determination of the value of 
$\sin ^2 \theta_{13}$ is quite good. It should be remembered 
that for values of $\sin ^2 \theta_{13} \le 10^{-6}$ there is 
no dependence on this variable as the H resonance is always 
non-adiabatic. 
Furthermore, the values of the $\nu_{\bar e}$ parameters are somewhat 
better determined than for the case A. 
The value of $\langle E_{\bar e} \rangle$ is determined 
within about 3 \% and $\beta_{\bar e}$ is constrained within 
about 8 \%. The electron-neutrino parameters 
are rather difficult to determine as there is not enough 
statistic in the $\nu_e$ detection channels. 
Nevertheless, in view of this, the allowed area of 
$\langle E_{e} \rangle$ is rather small,  
$\langle E_e \rangle$ is determined within 13\% but  
one can give only a lower limit for the luminosity 
$L_e>$ 0.1 $\times$ 10$^{32}$ ergs.

Let us next look at the contours for the case C in  
Fig.\ref{fig:cont}. These contours are in general very small. 
In particular, the degeneracy in the $\nu_x$ variables is 
broken down and the averaged $\nu_x$ energy is determined 
within 5 \%, which is a big improvement when compared to case B. 
Also the constraints on $\langle E_{\bar e} \rangle$ is somewhat 
improved, whereas the determination of $\theta_{13}$ and $\beta_{\bar e}$ 
have not really improved as compared to the case B. 
In contrast with the other cases, all the luminosities are constrained 
within about 10\% for the case C.

Of course the assumption of identical electron neutrino and 
electron antineutrino integrated luminosities can be criticized 
as it is not based on a strictly physical condition. Although the 
two luminosities are expected to be at least of the same order of magnitude, 
the ratio of the two is still varying a lot in different SN 
simulations. In Ref.\cite{sn1} this assumption was 
made~\footnote{In Ref.\cite{sn1} also the pinching parameters 
were fixed and the fiducial volume was 75\% smaller, {\it i.e.} smaller by a 
factor 540/720.} in order to decrease the  number of free 
parameters. Even if this assumption seems rather innocent 
we see that it has a large impact specially on the allowed region 
for the $\nu_x$ parameters. In fact, from Fig.\ref{fig:etafixed} it 
is clear, that either the assumption, $L_e=L_{\bar e}$ or the 
fixing of the $\beta_i$'s, significantly diminish the allowed 
average neutrino energies, most pronounced for $\langle E_x \rangle$. 
One can also observe, by looking at panels 3 and 5 of Fig.~\ref{fig:cont}, 
that the fixing of $\beta_x$ alone would allow one to determine quite well 
all the mean energies and luminosities (as long as the total luminosity is 
independently constrained).

However, neither the constraint $L_e=L_{\bar e}$ nor the fixing of the
pinching parameters, influence the allowed region for the Chooz angle
in a sizable way.  In the left panel of Fig.\ref{fig:etafixed}, we
have explicitly shown this by calculating the contour obtained when
fixing the $\beta_i$'s at their input values.  Therefore, a nice
feature that is kept even when including the pinching parameters is
the rather fine determination of the Chooz angle as seen in the last
panel of Fig.\ref{fig:cont}.  This is particularly important, since
for very small $\theta_{13}$, i.e. $\sin^2\theta_{13} \lsim 10^{-4}$,
the observation of a galactic supernova is, at least at present, the
only way of determining (or constraining) $\theta_{13}$.

Another important question that arises is whether the hierarchy 
can be determined by the observation of a galactic supernova? 
Here we consider simply the observation of the neutrino fluxes    
without the measurement of Earth matter or shock wave effects. 
Let us briefly comment on this point.
In this work we have fixed to neutrino mass hierarchy to normal.  But,
let us in the following discussion assume that the collective
neutrino-neutrino effects are negligible even for the inverted
neutrinos mass hierarchy.  In Ref.\cite{sn1} a discussion of the
prospects for determining the neutrino mass hierarchy is presented. 
In this work the $\beta_i$'s are fixed and also the condition
$L_e=L_{\bar e}$ has been implemented. Despite these assumptions it
seems clear to us that the results presented in Ref.\cite{sn1} about
the determination of the neutrino mass hierarchy will be valid even
for a 11 dimensional parameter space (i.e. that of
Eq.(\ref{paramspace}) along with the neutrino mass hierarchy).  The
hierarchy will influence the total number of events in each channel
for large values of $\sin ^2\theta_{13}$ (above $10^{-5}$) (see Table
II in Ref.\cite{sn1}).  Therefore the hierarchy can be determined by
the combined use of all four channels in a Water-Cherenkov detector.
The degeneracy that we have observed in the $\nu_x$ parameters on the
other hand maintains the total number of events in each channel almost
fixed and only changes the spectral form at low energies. Henceforth,
we expect that for large values of $\sin ^2\theta_{13}$, the neutrinos
mass hierarchy can be determined with very large statistic. Moreover,
in the case of the spectral swapping due to collective effects, as
described in Ref.\cite{collective}, it might even be possible to
determine the hierarchy for small values of $\sin ^2\theta_{13}$.

One can wonder to what extend our results depend on 
the particular input we have chosen. Especially if the 
average energies are further apart, like eg. 
$\langle E_x \rangle \gg \langle E_{\bar e} \rangle$, 
would the degeneracy in the $\nu_x$ parameters be much milder? 
Such investigation are left for a future paper.  

In the following we will discuss ways to break down the 
$\langle E_x \rangle$, $\beta_x$ and $L_x$ degeneracy. 
As the main problem is pinning down the 
value of $\langle E_x \rangle$, we will focus on the third frame 
in Fig.\ref{fig:cont}.
One possibility is that some robust features for the supernova 
parameters will emerge from supernova simulations in the future. 
These features can then be safely implemented in the fitting 
procedure and 
used to constrain the parameter space. This can be thought of 
as a standard supernova model, in analog to the use of the 
standard solar model when studying solar neutrinos.  
If a standard supernova model will be developed one can include 
a penalty in the $\chi^2$ function when eg. the ratio $L_e/L_{\bar e}$ 
differs from its central value as predicted by this standard model. 
We do not have such a standard model at our disposal yet, so to 
demonstrate our point we study some simple cases. 

Clearly, if the supernova simulations could tell us rather precisely
the value of the pinching parameters, then this degeneracy would be
broken.  It should be remembered that we fit to the total number of
events in the cooling phase. A slightly time varying spectral index,
will cause the time integrated spectra to be broader. Therefore, we
can expect lower values for the $\beta_i$'s in the time-integrated spectra.
By combining the information of the panels 3 and 5 in
Fig.\ref{fig:cont} one can directly relate constrains on $\beta_x$ to
constraints on $\langle E_x \rangle$.  Notice that if
we take the range value of 3--6, which is presently suggested by the SN
simulations, then the constraints on the average $\nu_x $ energy is
not improved for the case B.

\begin{figure}[htp]
\centering
\includegraphics[width=9cm,height=9.cm]{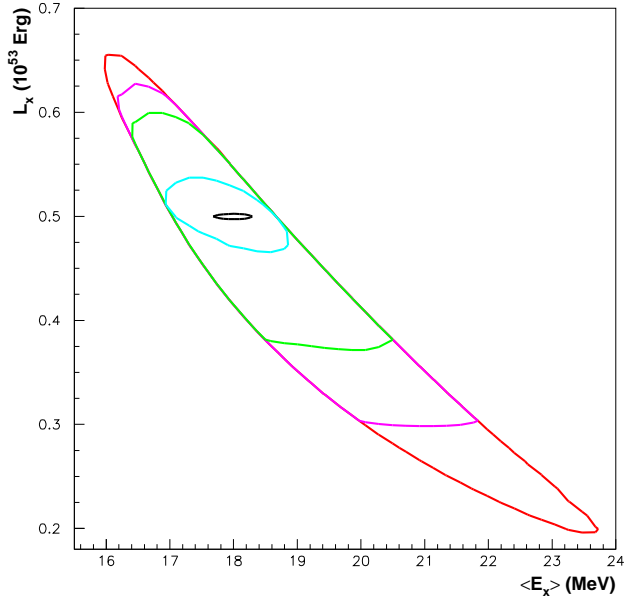}
\vspace{0.1cm}
\caption{The 3$\sigma$ allowed regions for various assumptions about 
constraints on the supernova luminosity ratios. We assume that all 
four channels are available in all the contours. The outermost dark 
(red) and the second smallest (cyan) contours are identical to the 
respectively case B and case C, in Fig.\ref{fig:cont} and 
shown for easy comparison. 
For the second largest (purple) contour we have constrained 
$0.5 \leq \xi_i \leq 2.0$ and for the third largest (green) contour 
$2/3 \; \leq \xi_i \leq 3/2$, where $i=1,2$. The smallest (black)
contour is calculated assuming $\xi_2=1$.
}
\label{fig:xi}
\end{figure}
An appealing possibility is to use ratios of the individual 
luminosities. Let us define two ratios 
\begin{equation}
  \xi_1 = \frac{L_e}{L_{\bar e}} \;\;\;,\;  
  \xi_2= \frac{L_x}{L_{\bar e}} \;.
\end{equation} 
One reason that these can be interesting parameters to use, is that
there is a good chance that these can be constrained (in a model
independent way) from supernova simulations.  The physical processes
that are behind the production of the neutrino flux are related to
known physics.  Nevertheless, there are a number of processes, such as
nucleon-nucleon bremsstrahlung ($NN \rightarrow NN \nu \bar \nu$),
neutrino-antineutrino annihilation (eg. $\nu_e \bar \nu_e \rightarrow
\nu \bar \nu$) and various scattering reactions between neutrino and
antineutrinos of different flavors, which of course complicates the
calculation of the neutrino emission. From the present supernova
simulations it seems that these ratios can at most be two and should be
larger than one-half.
It is important to notice that  when increasing the value of 
$\langle E_x \rangle$ within the degeneracy area, 
the ratios $\xi_1$ as well as $\xi_2$ also increases. 
In general, for values $\langle E_x \rangle \leq 18$ MeV the two ratios 
will be below one and for $\langle E_x \rangle \geq 18$
MeV the ratios will be larger than one. 
In Fig.\ref{fig:xi} we illustrate how the constrains on  
these ratios affect the allowed region. 
The second largest area in Fig.\ref{fig:xi} corresponds 
to the suggestions from present supernova simulations 
and a smaller allowed area is obtained.     
In Fig.\ref{fig:xi} we also show the contours when 
fixing $\xi_1=1$ (the same as the case C of Fig.\ref{fig:cont}) 
as well as $\xi_2=1$. In these cases a very good determination 
of $\langle E_x \rangle$ is possible. 
In conclusion we find that constraining the luminosity ratios 
can be helpful for pinning down the $\nu_x$ supernova parameters.

\begin{figure}[htp]
\centering
\includegraphics[width=9cm,height=9.cm]{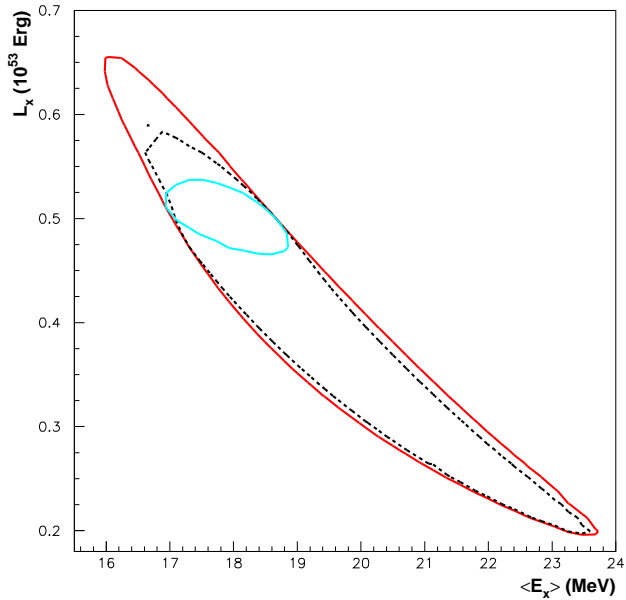}
\vspace{0.1cm}
\caption{The dashed dark (black) contour shows the 
3$\sigma$ allowed region when fixing the total luminosity 
of the supernova to the input value of $3\times 10^{53}$ ergs 
and leaving all  parameters free. We assume that all four 
channels are available. The outermost dark (red) and the 
light (cyan) contours are identical to the ones in 
Fig.\ref{fig:cont} and shown for 
easy comparison.
}
\label{fig:etot}
\end{figure}
Next, we will look at the possibility that the total 
energy liberated in the supernova, which we will refer to 
as $L_{\rm tot}$, has been already determined by some 
other method. 
In Ref.\cite{kachelriess} it is suggested 
that a measurement of the electron neutrinos from the 
neutronization burst can be used to determine the 
distance to the supernova. This along with a optical observation 
of the SN, might be used to predict the total energy liberated.  
In Fig.\ref{fig:etot} we present the results for fixing 
$L_{\rm tot}$ to its input value. This reveals that even in this 
extreme situation the improvement in the allowed region is minor.  
Indeed, the neutral currents channels, the NC-O and in parts 
also the ELAS channel, are already constraining the total luminosity. 
In case B the total luminosity is determined to about 4\% 
accuracy (assuming that the distance is known).  
Clearly constraining $L_{\rm tot}$ is not very helpful for 
determining the supernova parameters. In fact, one might 
even expect that the fitting procedure (case B) will provide the 
best determination of $L_{\rm tot}$.

\section{conclusion}
We have shown that the inclusion of all supernova and neutrinos
parameters is important for determining the allowed regions that can
be obtained from the observation of the neutrino burst from a
galactic supernova.  We use a total of 10 parameters ($\langle E_e
\rangle$, $\langle E_{\bar e} \rangle$, $\langle E_x \rangle$, $L_e$,
$L_{\bar e}$ $L_x$, $\beta_e$, $\beta_{\bar e}$, $\beta_x$ and $\sin^2
\theta_{13}$) and four detection channels (IB, CC-O, NC-O and ELAS)
that can be observed by a Water-Cherenkov detector to fit to such an
observation.  The degeneracy between $\langle E_x \rangle$, $\beta_x$
and $L_x$ when using only the inverse beta decay channel~\cite{mina},
is only broken mildly by the inclusion of the other channels in a
Water-Cherenkov detector. This is mainly due to the fact the number of
events from these other channels are at least one order of magnitude
smaller, and thus, in principle, the degeneracy could be broken by
including more channels if statistics was not a limitation.
Unfortunately, the supernova parameters are very difficult to
determine due to this degeneracy. Especially, the $\nu_x$ parameters
cannot be properly identified.  We have discussed ways that supernova
simulations can help overcoming this problem. A particular good way,
seems to be to constrain the ratio of the integrated luminosities of
the neutrino flavors.

We have demonstrated that the so-called Chooz angle, $\theta_{13}$,
can in principle be determined very well even when freely varying all
parameters (including the pinching parameters). In fact, whether or
not the pinching parameters are freely varied, does not influence much
the allowed $\theta_{13}$ region.  This is so because the Chooz angle
influence the ratios of the total number of events in each of the four
different channels, whereas the degeneracy caused by the pinching
parameters maintains the total number of events almost intact.  In the
same way we expect that the neutrino mass hierarchy can be determined
for large value of $\sin ^2 \theta_{13}$ by a galactic supernova.

On the other hand, data can only constrain the SN parameters 
describing the $\bar \nu_e$ flux ($\beta_{\bar e}$, 
$\langle E_{\bar e} \rangle$ and $L_{\bar e}$) and $\langle E_e \rangle$.
Without extra assumptions on the luminosities (ratios and/or total) or 
$\beta_x$ one cannot determine $L_e$, $L_x$ and $\langle E_x \rangle$, 
even if the four detection channels could be completely separated.

To summarize, the neutrino parameters can be determined quite
precisely, whereas it is more difficult to determine the supernova
parameters.
\begin{acknowledgments}
We would like to thank H. Minakata and H. Nunokawa for useful discussions 
and in particular we are grateful to HN for informing the results 
of the work in Ref.~\cite{mina} prior to the talk given by HM at 
NNN07~\cite{mina}.
SS thanks the Brazilian Funda\c{c}\~ao de Amparo \`a Pesquisa do 
Estado de S\~ao Paulo (FAPESP) for financial support during the 
first part of this work. SS is presently supported by a Nordita fellowship.
RZF thanks FAPESP and Conselho Nacional  de Ci\^encia e Tecnologia (CNPq) 
for partial support. 
\end{acknowledgments}

\end{document}